\documentclass{aastex63}
\usepackage{mathastext}
\graphicspath{{figures/}}
\usepackage{subcaption}
\usepackage{newfloat}
\usepackage{multirow}
\DeclareFloatingEnvironment[name={Supplementary Figure}]{suppfigure}
\usepackage{verbatim}
\usepackage{hyperref}
\hypersetup{draft}

\usepackage[section]{placeins}
\makeatletter
\AtBeginDocument{%
  \expandafter\renewcommand\expandafter\subsection\expandafter{%
    \expandafter\@fb@secFB\subsection
  }%
}
\makeatother

\received{--}
\revised{--}
\accepted{--}

\submitjournal{ApJ}

\begin{document}

\title{Chemical desorption \textit{versus} energy dissipation: insights from \textit{ab-initio} molecular dynamics of HCO formation}

\correspondingauthor{Stefano Pantaleone, Joan Enrique-Romero and Piero Ugliengo}
\email{stefano.pantaleone@univ-grenoble-alpes.fr}

\author[0000-0002-2457-1065]{Stefano Pantaleone}
\affiliation{Univ. Grenoble Alpes, CNRS, Institut de Plan\'{e}tologie et d'Astrophysique de Grenoble (IPAG), 38000 Grenoble, France}

\author[0000-0002-2147-7735]{Joan Enrique-Romero}
\affiliation{Univ. Grenoble Alpes, CNRS, Institut de Plan\'{e}tologie et d'Astrophysique de Grenoble (IPAG), 38000 Grenoble, France}
\affiliation{Departament de Qu\'{i}mica, Universitat Aut\`{o}noma de Barcelona, Bellaterra, 08193, Catalonia, Spain}
\email{juan.enrique-romero@univ-grenoble-alpes.fr}

\author[0000-0001-9664-6292]{Cecilia Ceccarelli}
\affiliation{Univ. Grenoble Alpes, CNRS, Institut de Plan\'{e}tologie et d'Astrophysique de Grenoble (IPAG), 38000 Grenoble, France}

\author[0000-0001-8886-9832]{Piero Ugliengo}
\affiliation{Dipartimento di Chimica and Nanostructured Interfaces and Surfaces (NIS) Centre, Universit\`{a} degli Studi di Torino, via P. Giuria 7, 10125, Torino, Italy.}
\email{piero.ugliengo@unito.it}

\author[0000-0001-5121-5683]{Nadia Balucani}
\affiliation{Univ. Grenoble Alpes, CNRS, Institut de Plan\'{e}tologie et d'Astrophysique de Grenoble (IPAG), 38000 Grenoble, France}
\affiliation{Dipartimento di Chimica, Biologia e Biotecnologie, Universit\`{a} di Perugia, Via Elce di Sotto 8, 06123 Perugia, Italy}
\affiliation{Osservatorio Astrosico di Arcetri, Largo E. Fermi 5, 50125 Firenze, Italy}

\author[0000-0002-9637-4554]{Albert Rimola}
\affiliation{Departament de Qu\'{i}mica, Universitat Aut\`{o}noma de Barcelona, Bellaterra, 08193, Catalonia, Spain}

\begin{abstract}
Molecular clouds are the cold regions of the Milky Way where stars form. They are enriched by rather complex molecules. Many of these molecules are believed to be synthesized on the icy surfaces of the interstellar submicron-sized dust grains that permeate the Galaxy.  At 10 K thermal desorption is inefficient and, therefore, why these molecules are found in the cold gas has tantalized astronomers for years. The assumption of the current models, called chemical desorption, is that the molecule formation energy released by the chemical reaction at the grain surface is partially absorbed by the grain and the remaining one causes the ejection of the newly formed molecule into the gas.
Here we report {an accurate} \textit{ab-initio} molecular dynamics simulations aimed to study the fate of the energy released by the first reaction of the H addition chain on CO, CO + H $\rightarrow$ HCO, occurring on a crystalline ice surface model. We show that about 90\% of the HCO formation energy is injected towards the ice in the first picosecond, leaving HCO with an energy content (10-15 kJ mol$^{-1}$) more than a factor two lower than its adsorption energy (30 kJ mol$^{-1}$). As a result, in agreement with laboratory experiments, we conclude that chemical desorption is inefficient for this {specific system, namely H + CO on crystalline ice}. We suspect this behavior to be quite general when dealing with hydrogen bonds, which are responsible of both the cohesive energy of the ice mantle and the interaction with adsorbates, as the HCO radical, {even though \textit{ad hoc} simulations are needed to draw specific conclusions on other systems}.
\end{abstract}



\section{Introduction}

Presently, more than 200 molecules are detected in the interstellar medium (ISM) \citep[e.g.][]{icom6}. Among them, all the molecules with more than 5 atoms contain carbon, the so-called interstellar Complex Organic Molecules \citep[iCOMs;][]{icom3,icom8}. 
iCOMs are especially observed in galactic star forming regions \citep[e.g.][]{icom1,blake_rotational_1986,icom2,icom5,lefloch_l1157-b1_2017,bianchi_astrochemistry_2019} and  external galaxies \citep[e.g.][]{muller_precise_2013,sewilo_detection_2018}. 
Besides, iCOMs are also detected towards cold ($\sim$10 K) sources \citep[e.g.][]{chem_des1,chem_des2,vastel_origin_2014,jimenez-serra_spatial_2016}. 
These last detections are important for (at least) two reasons: first, they challenge the idea that iCOMs are synthesized on the lukewarm (30-40 K) grain surfaces by radical-radical combination \citep{grain1,oberg_formation_2009,grain2}, and, second, if for whatever reason they are formed on the grain surfaces, the mechanism that lifts them off into the gas (where they are detected) must be non-thermal. 
Different non-thermal mechanisms have been invoked in the literature to explain the presence of gaseous iCOMs in cold environments: cosmic-ray spot heating \citep{leger_desorption_1985,hasegawa_new_1993}) or sputtering \citep[e.g.][]{dartois_non-thermal_2019}, UV-induced photo-desorption \citep[e.g.][]{dominik_gas-phase_2005,fayolle_co_2011,bertin_indirect_2013,bertin_uv_2016}, co-desorption of ices \citep[e.g.][]{sandford_condensation_1988,abund3}, and chemical (or reactive) desorption {\citep{DW1993,garrod_non-thermal_2007,minissale_influence_2014,non_chem_des}}.

Here, we focus on the last mechanism, the chemical desorption (CD). The underlying idea is that the energy released by strongly exothermic chemical reactions occurring on the grain surfaces is only in part absorbed by the grain while the remaining one is used to break the bonds of the newly formed species with the surface, so that a fraction of the synthesized species is injected into the gas phase. Therefore, CD and the dissipation of the surface-reaction energy are two faces of the same medal, intrinsically linked.

From an experimental point of view, it is extremely difficult, if not impossible, to quantify the energy dissipation. Overall, laboratory experiments showed that CD can be more or less efficient depending on the adsorbate and the substrate. For example, \cite{oba_infrared_2018} found high CD efficiencies {($\sim$60\%)} for H$_2$S formation on amorphous solid water. {On the contrary, lower CD efficiencies for different systems were observed by other authors: \cite{dulieu2013} found a CD efficiency for the O$_2$ + D reaction lower than 10\%, \cite{He_2017} showed that the H addition to O$_3$ causes the desorption of the product O$_2$ by no more than 11\%, and \cite{Chuang_2018} found a CD efficiency lower than 2\% per H-atom induced reaction in the hydrogenation of CO towards methanol.} In a systematic study of CD in several reactions on different substrates, \cite{non_chem_des} showed that the CD efficiency {does indeed depend} on three major factors: the reaction formation energy, the binding energy of the adsorbate, and the {nature} of the substrate. They proposed a general formalism to estimate the CD probability, based on the idea that the energy dissipation can be approximately treated as an elastic collision.

Theoretical calculations are, in principle, capable to simultaneously study the energy dissipation and CD. Various techniques have been so far used for different systems. \cite{mol_dissip1} and \cite{mol_dissip2} simulated the relaxation of translationally excited admolecules (CO$_2$, H$_2$O and CH$_4$) on crystalline and amorphous ice models, via classical Molecular Dynamics (MD) simulations. 
They run thousands of simulations using approximate interaction potentials where the admolecules were given large (0.5 to 5 eV, equal to 50 to 500 kJ mol$^{-1}$) translational energies in random directions. Therefore, 
in these studies, all the chemical reaction energy is assumed to be canalized into translational motion of the admolecules {and, for this reason, a statistical analysis was necessary to understand the fate of the energy for different injected trajectories}. Thus, each simulation follows the evolution of the energy of the species recording whether the molecule is desorbed, penetrates in the surface or is adsorbed on the surface. Based on their large number of simulations, \cite{mol_dissip1} and \cite{mol_dissip2} found that the desorption probability depends on the injected kinetic energy and binding energy of the species. Additionally, they provided a formula to estimate the CD probability, which depends on those two quantities. Despite the careful and exhaustive statistical analysis on the simulations, there are some weaknesses to be pointed out:
(i) the limits of force field-based methods in dealing with chemical reactions;
(ii) the injected energy of the admolecule which is only translational, while, after a reaction occurs, the energy should be partitioned also into vibrational and rotational levels;
(iii) the relatively small size of the used water clusters, whose temperature may rise for high energy injections ($\sim$5 eV), thus resulting in an overestimation of the admolecule mobility.

\cite{korchagina_molecular_2017} studied the energy dissipation of the hydrogenation reaction of CO (producing HCO) under the Eley-Rideal model at temperatures of 70 K. They used MD simulations at the self-consistent-charge density functional tight binding (SCC-DFTB) theory level, which is an approximation of classical DFT, based on parametrized integrals and charges. They used small molecular clusters, from 1 to 10 water molecules, to simulate the ice surfaces and showed that the energy released after HCO formation can be dissipated (i.e. the reaction gives a stable HCO) on clusters with N$\geq$1 water molecules, and that the product remains adsorbed on the clusters (i.e. no CD) for clusters with N$\geq$3. Such behavior is linked to the capacity of water to redistribute the reaction energy excess into vibrational excitation. However, such small clusters cannot represent the ice mantle when dynamical effects are taken into account, because, although very small clusters can stabilize the product by absorbing efficiently the nascent energy, the mobility of the molecule is strongly overestimated.
Indeed, the  cluster’s temperature increase linearly depends on the water molecules number of the surface model: it should be large enough to be able to harness the reaction energy while restraining the temperature increase.

{\cite{kayanuma2019} studied the reaction of H with adsorbed HCO on a graphene surface by means of \textit{ab-initio} Molecular Dynamics (AIMD) simulations, showing that in the case of HCO chemisorption (i.e. chemical bond between the adsorbate and the surface), the products H$_2$ + CO are desorbed, while in the case of HCO physisorption (the interaction with the surface is of dispersive nature), formaldehyde is formed without chemical desorption.}

Here we present new AIMD simulations of the reaction H + CO on a large periodic crystalline water-ice surface assuming the Langmuir-Hinshelwood mechanism. 
{This reaction is the first step towards the formation of methanol on the grain surfaces, one of the most studied both theoretically and experimentally \citep[e.g.][]{Watanabe2002, Woon_2002, Hiraoka2002, Watanabe2007, hco_tunnel, Rimola2014}; therefore, in this context, it can be considered as one of the most important reactions in astrochemical studies. In addition, it is representative of the class of reactions with a relatively low reaction energy (less than 2 eV) to dissipate.}
Our scope is to understand from an atomic point of view how the energy released by the HCO formation is transferred towards the water surface, without any \textit{a priori} assumption on how the reaction energy is distributed over the system. 
{We emphasize that our approach is substantially different from the one used by \cite{mol_dissip1} and \cite{mol_dissip2}, described above. In our case, we do not need to address a statistical behavior (depending on the species trajectory), because we simulate the reaction itself and how its energy is dissipated by the formed HCO. This does not depend on the initial H trajectory because the energy of the H atom is thermal (at 10 K, specifically) and, therefore, negligible with respect to the energy released by the reaction (about 1.4 eV). The result could, in principle, depend on the initial position of the CO on the crystalline ice, for which there are a few possibilities, and we will discuss this point in the article. In summary, our AIMD simulations allow us to quantify whether} the newly formed species has enough energy to break its interactions with the water surface and, consequently, to be injected into the gas-phase.

{Last,} despite it is well known that interstellar ice is {often} amorphous, we chose a crystalline model because tuning the computational setup is easier. Once the system is carefully tested, our future works will focus on amorphous ice models. 
{It is important to notice, however, that crystalline water ice has been detected in the ISM \citep{Molinari_1999} and, particular relevant for the planet formation studies, in protoplanetary disks \citep{terada2012}, so that our simulations will be directly applicable in those environments.}

The {article} is organized as follows. In Sec. \ref{comp} we present the computational methodology, in Sec. \ref{res} the results and in Sec. \ref{disc} we discuss these results in view of astrochemical implications. {Finally, in Sec. \ref{concl}, we summarize the most important conclusions.}

\section{Computational Details}
\label{comp}

\subsection{Methods}

All the calculations have been carried out with the CP2K package \citep{cp2k1, cp2k3, cp2k6, gpw, pseudo1, pseudo2}. The atoms have been treated as follows: core electrons have been described with the Goedecker-Teter-Hutter pseudopotentials \citep{pseudo1,pseudo2}, while valence electrons with a mixed Gaussian and Plane Wave (GPW) approach \citep{gpw}. 
The PBE functional has been used for all the calculations \citep{pbe} combined with a triple-$\zeta$ basis set for valence electrons plus 2 polarization functions (TZV2P). 
The cutoff for plane waves has been set to 600 Ry. 
The \textit{a posteriori} D3 Grimme correction has been applied to the PBE functional to account for dispersion forces \citep{grimme1,grimme2}. 
During the optimization procedure, only the H, C, and O atoms (the ones belonging to the HCO$\cdot$) were free to move, while the atoms belonging to the ice surface have been kept fixed to their thermalised positions.All calculations were carried out within the unrestricted formalism as we deal with open-shell systems. 
The spin density was checked for reactive, TS and product and it remains always well localized either on H atom (reactant) and on HCO for the product. No spread of spin density through the ice was detected ((see Supplementary Figures 2, 3, and 4)). 
The binding energy (BE) of HCO$\cdot$ were calculated according to 
$BE_{HCO}=E_{CPLX}-(E_{Ice} + E_{HCO})$.


Where $E_{CPLX}$ is the energy of the HCO/Ice system, $E_{Ice}$ that of the bare ice surface, and $E_{HCO}$ the energy of the HCO$\cdot$ alone, each one optimized at its own minimum. The BE$_{HCO}$ will be used later on to compare with the residual kinetic energy of the HCO formation.

In order to reproduce the ISM conditions, the reaction was carried out in the microcanonical ensemble (NVE), where the total energy (\textit{i.e.} potential + kinetic) is conserved. Moreover, we run an equilibration AIMD in the NVT ensemble (using the CSVR thermostat, with a time constant of 20 femtoseconds) at 10 K for 1 ps (with a time step of 1 fs) for the bare ice surface, to obtain a thermally equilibrated ice. Accordingly, the equilibrated velocities of the ice surface were used as starting ones for the NVE production, while the H and C velocities of HCO were manually set according to the H--C bond formation. The evolution of the system was followed for 20 ps, using a timestep of 1 fs.

{In addition, in the Annex, we present results obtained from a benchmark study on the reaction of HCO formation on a small cluster of 3 H$_2$O molecules (H + CO/3H$_2$O $\longrightarrow$ HCO/3H$_2$O), similarly to the work by \cite{Rimola2014}. Results shows that PBE overestimates the energetics of the reaction (132 kJ mol$^{-1}$ vs 91 kj mol$^{-1}$, provided by the CCSD(T) method.) By contrast, frequency calculations, which are responsible of the vibrational coupling of the water molecules with the HCO and, accordingly, of the kinetic energy dissipation efficiency, are in good agreement to those calculated at higher level of theory.}

\subsection{Ice model}

Ordinary ice is proton disordered and, accordingly, its crystal structure cannot be simply modeled by adopting relatively small unit cells. A possible alternative is to adopt P-ice, a proton ordered ice already successfully used in the past to simulate ice features \citep{ice}. P-ice bulk belongs to the Pna2$_1$ space group and from the bulk we cut out a slab to simulate the (100) surface, shown in Supplementary Figure 5. The size of the surface was chosen according to the amount of energy to be dissipated. Given that the HCO$\cdot$ radical formation is strongly exothermic (132.5 kJ mol$^{-1}$; see Figure \ref{hco_prod}), a sufficiently large water ice slab is needed to absorb most of the nascent energy (see more details in the next section). Therefore, the periodic cell parameters have been set to $\textit{a} = 17.544$ \r{A} and $\textit{b} = 21.2475$ \r{A} with a slab thickness of $\sim13$ \r{A} (which corresponds to 4 water layers). The model consists of 192 water molecules in total. In the CP2K code, the electron density is described by plane waves, and, accordingly, the surface is replicated also along the non-periodic direction. To avoid interactions between the fictitious slab replicas, the \textit{c} parameter (\textit{i.e.}, the non-periodic one) was set to 35 \r{A}.

\section{Results}
\label{res}

\begin{figure*}
\centering
	\begin{subfigure}[b]{0.3\textwidth}
	\includegraphics[width=\textwidth]{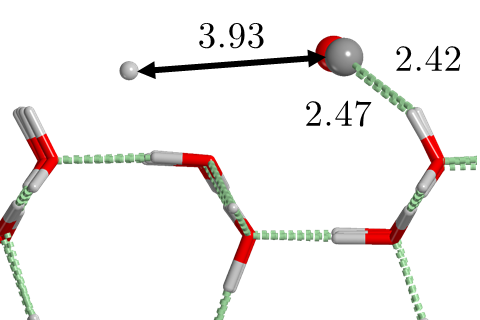}
	\caption{Reactant (0.0)}
	\label{hco_reag}
	\end{subfigure}
	\hspace{5pt}
	\begin{subfigure}[b]{0.3\textwidth}
	\includegraphics[width=\textwidth]{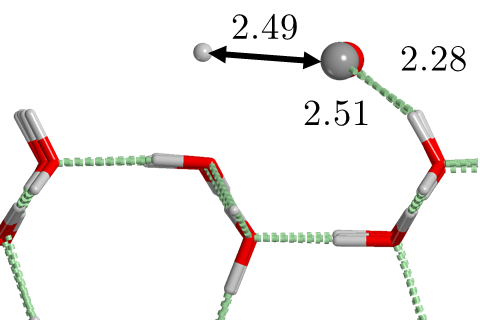}
	\caption{Transition state (5.2)}
	\label{hco_ts}
	\end{subfigure}
	\hspace{5pt}
	\begin{subfigure}[b]{0.3\textwidth}
	\includegraphics[width=\textwidth]{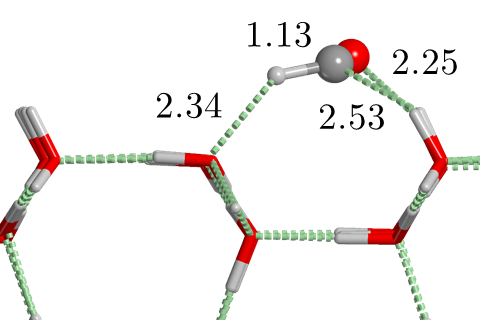}
	\caption{Product (-132.5)}
	\label{hco_prod}
	\end{subfigure}
\caption{PBE-D3 optimized geometries of reactant, transition state and product of the HCO$\cdot$ reaction formation on the ice surface. The numbers in parenthesis correspond to the relative energy in kJ mol$^{-1}$ with respect to the reactant. Distances are in \r{A}. H atoms in white, C atom in gray, O atoms in red.}
\label{reaction}
\end{figure*}

To study the CO hydrogenation on the ice surface, we simulated the reaction adopting a Langmuir-Hinshelwood {(LH)} surface mechanism, \textit{i.e.} with both the reactants (H$\cdot$ and CO) adsorbed on the surface. Accordingly, we first optimized the geometries of the reactants (H$\cdot$ + CO), transition state (H$\cdots$CO) and product (HCO$\cdot$) in order to obtain the potential energy surface of the reaction. As reported in Figure \ref{reaction}, the activation barrier (5.2 kJ mol$^{-1}$, 622.1 K) is quite high if we consider the sources of energy available in the ISM. Indeed, it is well known that this reaction proceeds mostly through H-tunneling (\cite{Hiraoka2002,hco_tunnel,Rimola2014}). However, as AIMD operates within the Born-Oppenheimer approximation, \textit{i.e.} the nuclei motion is driven by classical equations, quantum phenomena of atoms (such as tunneling effect) cannot be taken into account. Since our aim is not to simulate the reaction itself, but to understand where does the liberated energy go, we run the simulation starting from the transition state structure (Figure \ref{hco_ts}). In this way, we force the system to evolve in the direction of the product. Therefore, the total energy to be dissipated is the sum of the energy barrier and of the reaction energy (5.2 + 132.5) kJ mol$^{-1}$ = 137.7 kJ mol$^{-1}$. Therefore, it is possible to estimate the expected temperature increase of the whole system after the reaction by invoking the equipartition theorem:

\begin{equation}
T = \frac{2}{3}\frac{E_{nasc}}{R~N_{at}}
\end{equation}

where E$_{nasc}$ is the nascent energy due to the H--C bond formation (\textit{i.e.} 137.7 kJ mol$^{-1}$), N$_{at}$ is the whole system atoms number (3 for HCO$\cdot$ and 3 $\times$ 192 for the ice), and R the gas constant. Thus, the energy dissipation trough a ice slab containing 192 water molecules should produce a global temperature increase of about 19 K (which is in perfect agreement with the very first spike in T of Supplementary Figure 7, reaching 29 K = (10K + 19K) where 10K is the starting temperature). Then, when the simulation equilibrates, the temperature oscillates around 20 K. This very simple calculation is useful to have an idea of the atom number needed to avoid the nascent energy to artificially rise the total temperature.

\begin{figure}
\centering
\includegraphics[width=0.49\textwidth]{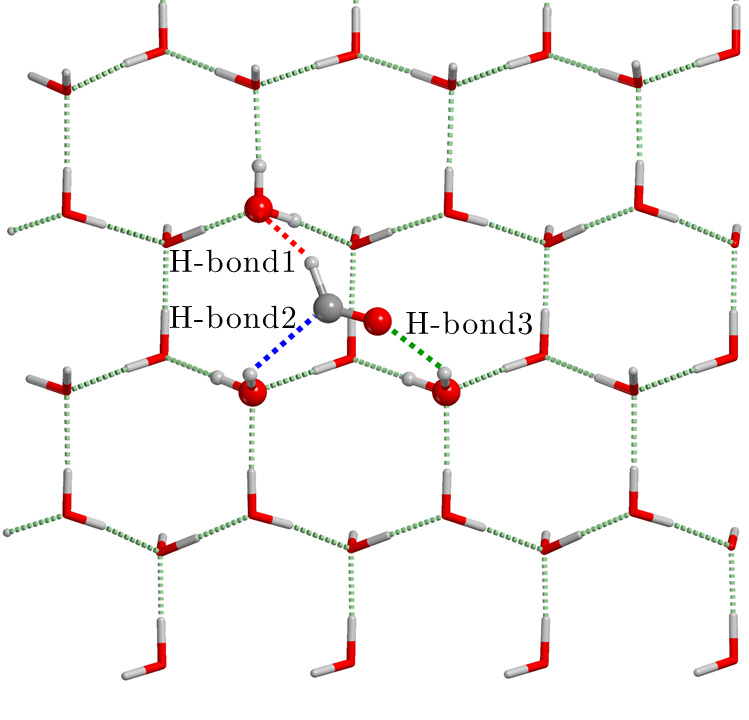}
\includegraphics[width=0.49\textwidth]{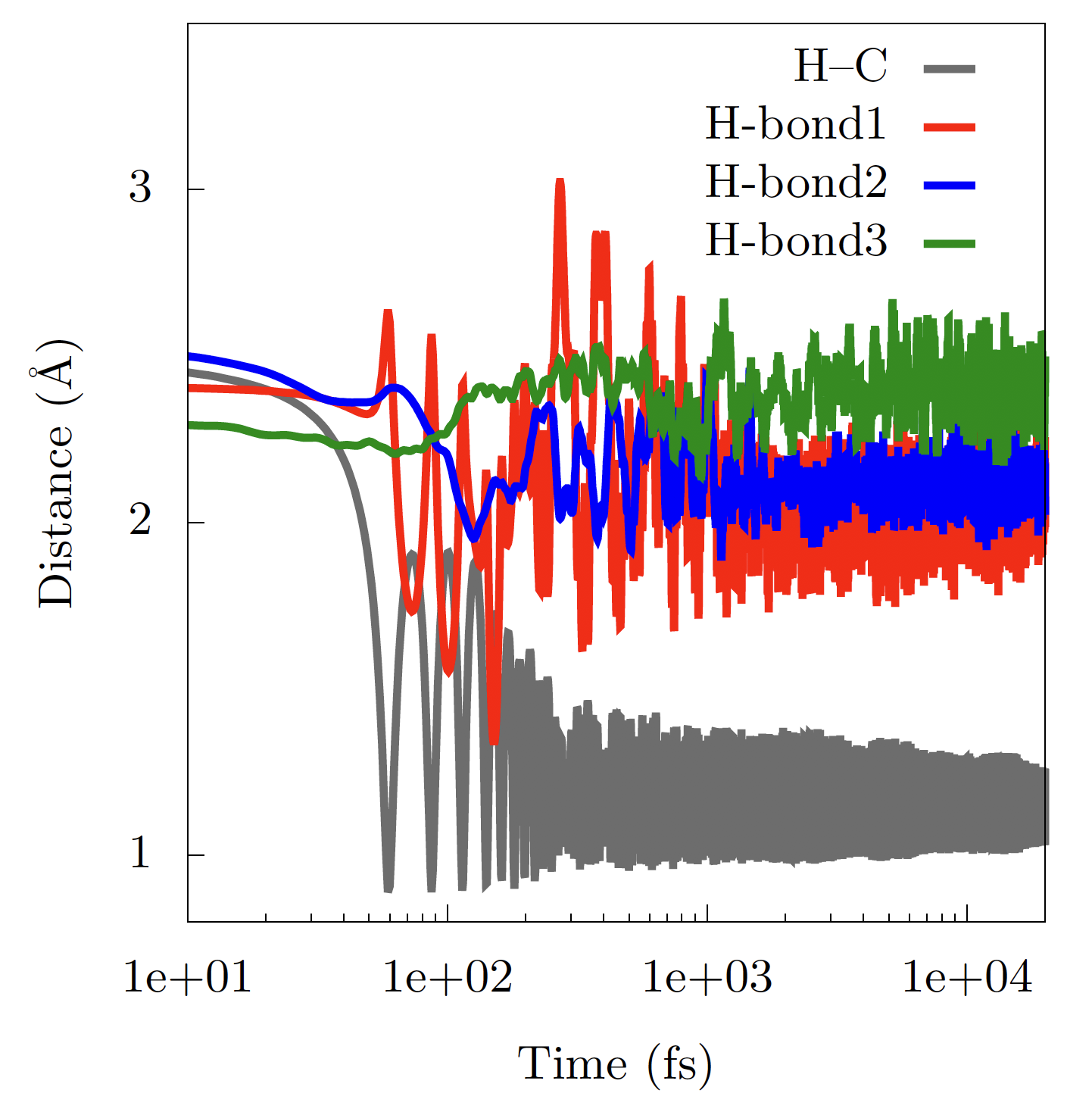}
\caption{Structure of the HCO$\cdot$ adsorbed on the ice surface at the last point of the AIMD simulation (left) and evolution of the most relevant geometrical parameters during the AIMD simulation (right). H-bonds colors in the chart correspond to the H-bonds in the figure depicted as dotted lines. H atoms in white, C atom in gray, O atoms in red.}
\label{evol}
\end{figure}

Figure \ref{evol} shows the most interesting geometrical parameters of the system during the AIMD simulation. Both the temperature and potential energy oscillate around a stable value and they reach the equilibrium within 1 ps (see also Supplementary Figures 7 and 8). As one can see in Figure \ref{evol}, the H--C bond forms in the first tens of fs of the simulation (keeping in mind that we are starting from the transition state). After the H--C bond formation, the HCO$\cdot$ radical moves on the surface in the sense of maximizing the H-bond contacts with the surrounding water molecules (Figure \ref{evol}) and it lies in its most stable position after 1 ps, which corresponds to the equilibration of both the potential energy and the temperature. After this period, the HCO$\cdot$ stays in this stable position, without diffusing anywhere.

\begin{figure*}
\centering
\includegraphics[width=0.7\textwidth]{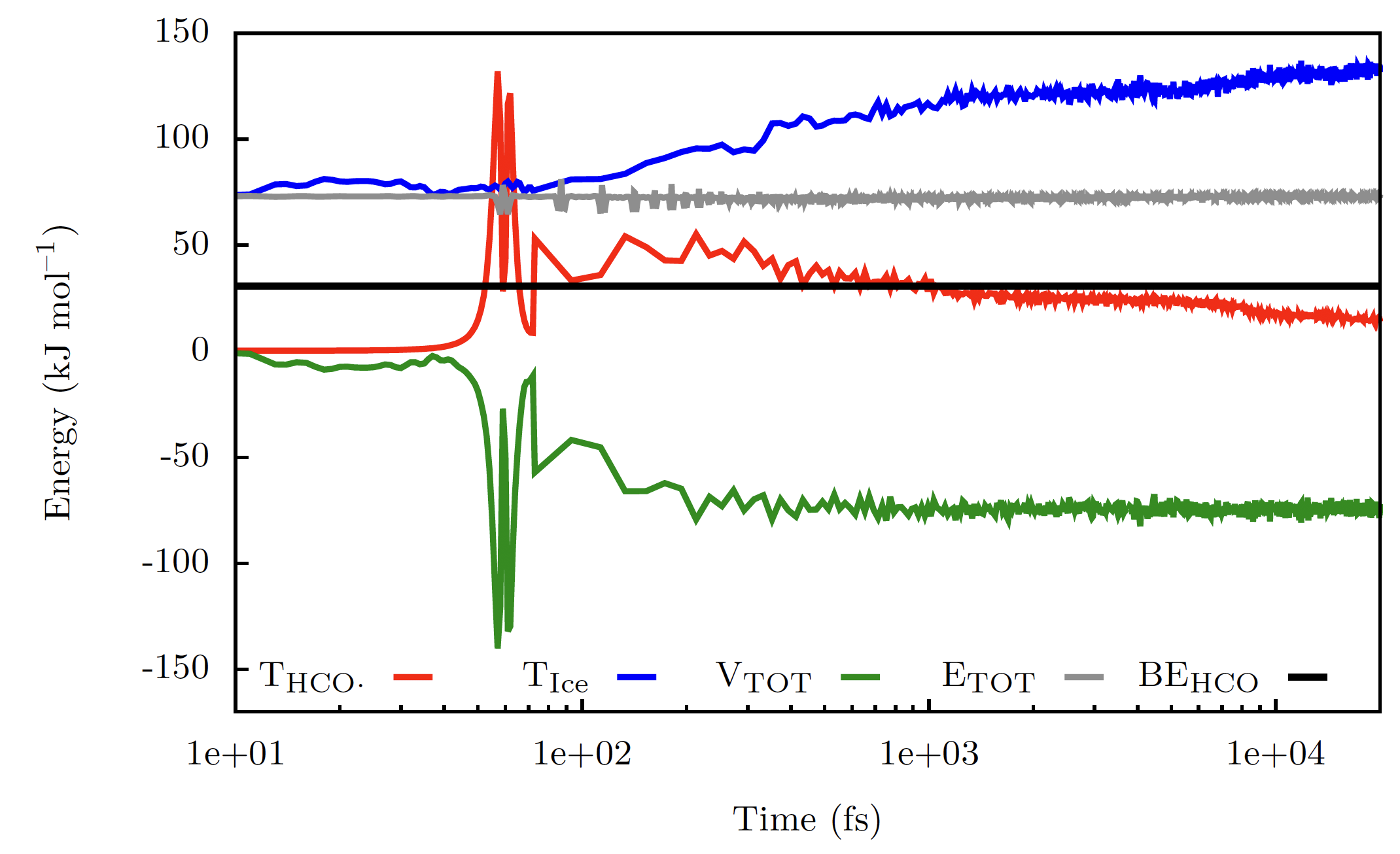}
\caption{Evolution of the most relevant energetic components (in kJ mol$^{-1}$) of the HCO$\cdot$/Ice system during the AIMD simulation. E$_{TOT}$ is the total energy (\textit{i.e.} potential + kinetic, gray line). V$_{TOT}$ is the potential energy (green line). T$_{HCO\cdot}$ and T$_{Ice}$ are the kinetic energies of HCO$\cdot$ (red line) and ice (blue line), respectively. BE$_{HCO\cdot}$ is the binding energy of the HCO$\cdot$ (black line). Gray line shows very good energy conservation.}
\label{tot_kin}
\end{figure*}

In Figure \ref{tot_kin}, the kinetic energy dissipation due to the H--C bond formation is reported. As expected, the kinetic energy released from the H--C bond formation rapidly drops (in less than 100 fs) and it is, simultaneously, absorbed by the water molecules of the surface. As one can see, at the beginning of the simulation, before 200 fs, the red line (T$_{HCO}$) is complementary to the green line (V$_{TOT}$). As the AIMD was executed in the NVE ensemble, when the kinetic energy drops, the potential one rises by the same amount. However, after 300 fs, when the HCO$\cdot$ starts to exchange energy with the surface, the red (T$_{HCO}$) and blue (T$_{Ice}$) lines become symmetric, \textit{i.e.} the energy loss of HCO$\cdot$ is equal to the energy gain of the surface, in terms of kinetic energy. The most important message from Figure \ref{tot_kin} is that, within the first ps, HCO$\cdot$ looses 90\% of its initial kinetic energy, which is immediately transferred to the ice. Later, along the simulation, HCO$\cdot$ continues to transfer energy to the ice at a slower rate. After 20 ps, its kinetic energy is around 15 kJ mol$^{-1}$, at least twice lower than its binding energy. 
Therefore, it is unlikely that the HCO$\cdot$ will desorb. This is corroborated by Figure \ref{evol}, where the H-bonds of HCO$\cdot$ to the surface lie in a rather steady fashion after its formation (they essentially vibrate around their equilibrium position).

In order to give a detailed analysis of the energy dissipation on the ice surface, we used the atomic simulation environment (ASE) python module \citep{ase1,ase2}. The energy {dissipation} was analyzed by dividing the {slab of} water {molecules} in concentric shells {centered on the reaction center (\textit{i.e.} the C atom). 
Note that the HCO radical has been excluded from this analysis because we are interested in the dissipation across the ice itself. The concentric shells were defined as follows: i) the first one is a hemi-sphere with 4 \r{A} of radius and contains the closest water molecules to the CO molecule; ii) the other shells are equally spaced from each other by 2.8 \r{A} (average closest O--O distance between water molecules), up to a distance of 18 \r{A} in order to include also the farthest water molecules from the reaction center. 
We emphasize that only a single unit cell was used for this analysis, which means that no water molecules from periodic replicas are included in the energy dissipation analysis. The sketch showing the water shells is reported in Supplementary Figure 11.}

\begin{figure*}
\centering
\includegraphics[width=0.7\textwidth]{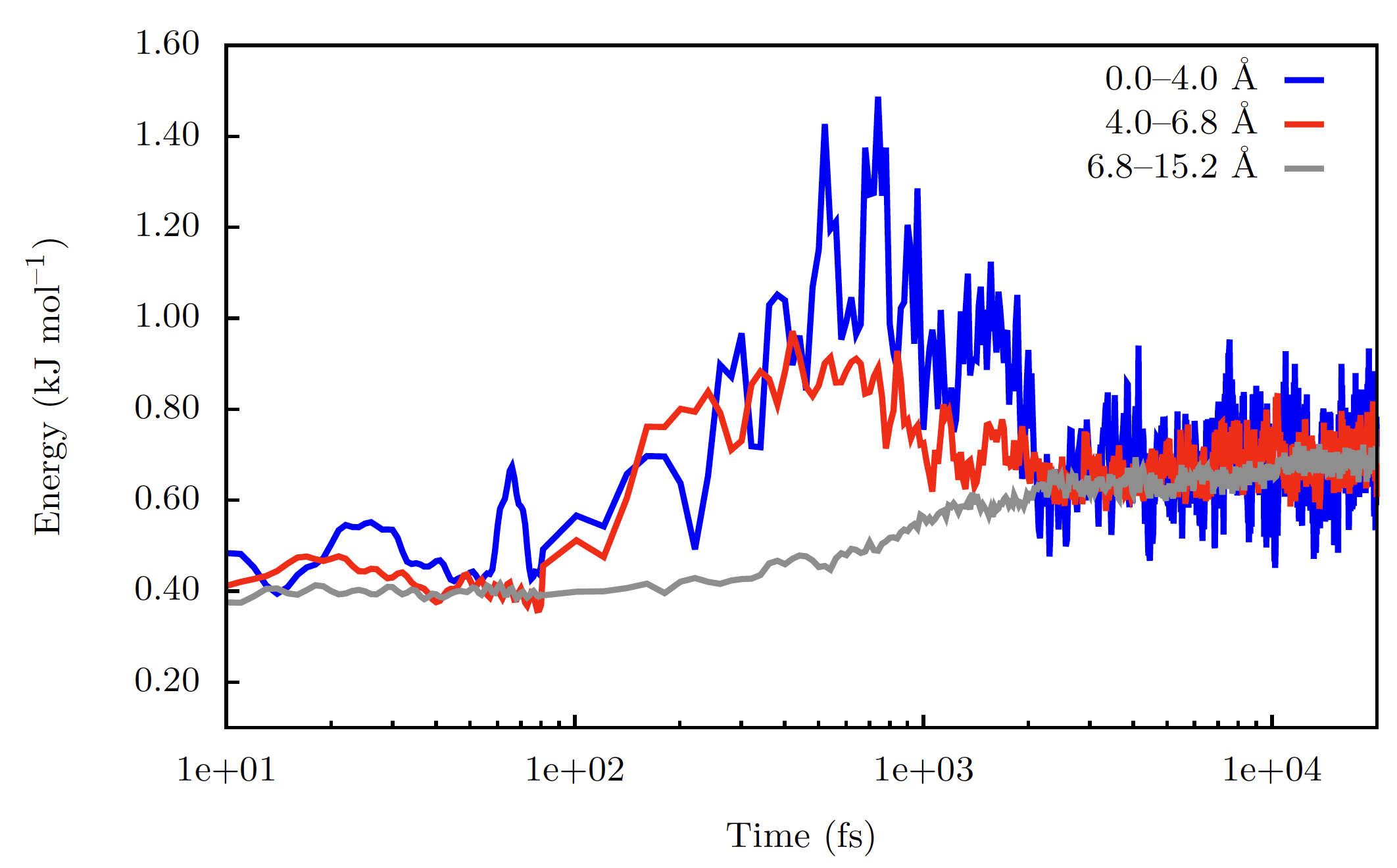}
\caption{Kinetic energy evolution (in kJ mol$^{-1}$ per water molecule) of the ice surface. The ranges in the legend refer to the shells the water molecules belong to.}
\label{wat_kin}
\end{figure*}

{The results on the energy dissipation analysis} are shown in Figure \ref{wat_kin}. The kinetic energy was normalized per water molecule, in order to remove the dependence on the number of water molecules (as each shell contains different number of waters). 
One can see, from the two first shells {(0.0-4.0 and 4.0-6.8 \r{A})}, that the energy is rapidly transferred from HCO to the ice, which is later uniformly distributed to outer shells: within $\sim$2 ps all shells have the same kinetic energy.

\section{Discussion}
\label{disc}

In the present work, the HCO$\cdot$ formation reaction on an ice surface model has been used as a test case in order to study the kinetic energy dissipation due to a bond formation (in the specific case the H--C bond). This is particularly important because the energy released by the formation of a chemical bond may be a hot spot that makes possible other processes, like the desorption of the newly formed molecule or other molecules nearby. A crucial parameter is the timescale of the process: the main question is whether the released energy is available (and if yes, for how long) and {whether it} can be used by other adsorbed species, or, in contrast, it is immediately {dissipated} through the ice surface. From Figure \ref{tot_kin}, the answer is very clear: the slab of water molecules absorbs $\sim 90 \%$ of nascent energy, which is equally distributed among all the water molecules of the ice surface within the first ps after the bond formation, and it is no longer available to assist other processes.
In particular, and importantly, the {residual HCO} kinetic energy after 1 ps is almost half of the HCO binding energy, which implies that HCO will remain stuck on the surface {and will not desorb}.

{Our simulations are based on three assumptions: (i) the starting position of the CO adsorbed on the ice is the most energetically favorable one; (ii) the surface of the ice is crystalline; (iii) the reaction follows a LH mechanism. 
In the following, we discuss the validity limits of each assumption.}

{(i) CO position:
The first assumption is based on the fact that, as molecules in cold molecular clouds move at low, thermal ($\sim$10 K) velocities, landing on grain surfaces is slow enough for them to feel the electrostatic potential generated by the surface. Consequently, they have sufficient time to accommodate on the icy mantle, maximizing their interactions with the ice surface itself. In other words, the main driving forces of the adsorption process in cold molecular clouds are long range forces. 
Nonetheless, a few other starting positions may exist with respect to the one that was chosen for this study.
For this reason, we have explored two other starting configurations, namely the Pos2 and Pos3 reported in Supplementary Figure 6.
Both of them, after either geometry optimisation (Supplementary Figures 6a and 6b) or AIMD simulation (Supplementary Figures 6c and 6d), lead to the HCO radical in the same position of that reported in Figures \ref{hco_prod} and \ref{evol}.
In other words, whether CO is in the position we chose for the full AIMD simulations or in the other two positions, HCO ends up having the same position, which means the same bonds with the surface and, consequently, the evolution of the system is the same.}

{(ii) Crystalline ice:
As already mentioned in the Introduction, the major reason for choosing the crystalline ice structure is a computational one.}
{In this respect,} we would like to caution about the role of crystalline versus amorphous ice, because of the possibility that the symmetric {electrostatic potential of the crystalline case} can hinder the formed species to escape from the surface {and, in crystalline systems, the vibrational coupling might be more efficient than in amorphous ones, thus allowing a faster dissipation of the energy and, consequently, underestimating the desorption rate.}
Further studies need to be carried out in order to understand if the crystalline nature of the ice affects and how the behavior of the formed species because of the crystal symmetry compared to the random nature of the amorphous ice.
{Having said that, our simulations are valid and applicable in the environments where crystalline ice has been detected \citep[e.g.][]{Molinari_1999,terada2012}.}

{(iii) LH mechanism:
It is possible, and even probable, that the H atom will not arrive directly from the gas but rather is an atom that randomly grazes the ice surface. In this case, since the velocity of the H is even smaller than the one if it landed from the gas, the results of our simulations would not change. So, this choice is, after all, irrelevant for the purpose of our study.}

{In summary}, we conclude that chemical desorption is not efficient in the H + CO reaction {on crystalline ice, and this is a robust result.} 
{Laboratory experiments have proven to be difficult to obtain for this specific reaction. 
To our best knowledge, no experiment simulates it on crystalline ices. 
\cite{non_chem_des} obtained a measure of the H + CO CD when the reaction occurs on oxidized graphite. They found a CD efficiency equal to 10$\pm$8 percent. 
\cite{Chuang_2018} studied the CO hydrogenation process using as substrate gold, over which CO was deposited forming a thick layer of solid CO, subsequently bombarded with H atoms. 
They found that the global CD efficiency of the whole process up to the formation oh CH$_3$OH is low, $\leq$0.07 per hydrogenation step, assuming an identical efficiency for each reaction in the hydrogenation process \citep{Chuang_2018}.
As already mentioned, the surface where the CO is adsorbed and the reaction occurs is certainly of paramount importance, so that it is not obvious to make a direct comparison between our computations and the above experiments. Yet, it seems that then latter agree on a small CD efficiency, if any, as our computations predict.}

Finally, it is possible that for more exothermic reactions (like for example the last step to CH$_3$OH, which is much more exothermic than H + CO) and  weakly bound systems (like H$_2$), chemical desorption {can take place}. 
This could also be the case for reactions occurring on grain surfaces of different nature, such as silicates or carbonaceous materials, as their heat capacities are very different from those of H$_2$O-dominated ices.
Dedicated simulations should be carried out to assess it in these systems.

\section{Conclusions}
\label{concl}

{We studied the first step of the hydrogenation of CO on the interstellar grain icy surfaces by means of ab-initio Molecular Dynamics simulation. We studied the H + CO reaction occurring on a crystalline ice. Our goal was to understand from an atomistic point of view and to quantify the possibility that the energy released in the reaction is just partially dispersed on the crystalline substrate and the residual one is used to desorb the product, HCO.}

{The main conclusions of the present study are:\\
(i) The reaction energy dissipation through thermal excitation of water molecules is extremely fast: after the first picosecond most of the reaction energy (137.7 kJ mol$^{-1}$) is dissipated away through the ice, leaving HCO with a kinetic energy of 10--15 kJ mol$^{-1}$, more than twice lower than its binding energy (30 kJ mol$^{-1}$).\\
(ii)As a consequence, the HCO product is doomed to remain attached to the crystalline ice and no desoprtion can occur.}

{The astrophysical implications are that, in the environments where crystalline ices are present, like for example some protoplanetary disks, chemical desorption does not occur for the reaction H + CO.
We suspect that this may be a general behavior for reactions dealing with hydrogen bonds, as they are responsible for both the cohesive energy and the interaction with the crystalline ice. However, in order to assess whether this is true, ad hoc simulations similar to those presented here are mandatory.}

\section{Acknowledgements}
The authors acknowledge funding from European Union's Horizon 2020 research and innovation program, the European Research Council (ERC) Project ``the Dawn of Organic Chemistry", grant agreement No 741002, and the Marie Skłodowska-Curie  project ``Astro-Chemical Origins” (ACO), grant agreement No 811312.
PU and NB acknowledge MIUR (Ministero dell' Istruzione, dell' Università e della Ricerca) and Scuola Normale Superiore (project PRIN 2015, STARS in the CAOS - Simulation Tools for Astrochemical Reactivity and Spectroscopy in the Cyberinfrastructure for Astrochemical Organic Species, cod. 2015F59J3R).
AR is indebted to the ``Ram{\'o}n y Cajal" program. MINECO (project CTQ2017-89132-P) and DIUE (project 2017SGR1323) are acknowledged.
BSC-MN and OCCIGEN HPCs are kindly acknowledged for the generous allowance of supercomputing time through the QS-2019-2-0028 and 2019-A0060810797 projects, respectively.

\end{document}